\documentstyle[aps]{revtex}

\newcommand{\BE}{\begin{equation}}
\newcommand{\EE}{\end{equation}}
\newcommand{\BA}{\begin{eqnarray}}
\newcommand{\EA}{\end{eqnarray}}

\begin{document}

\title{A phase transition in acoustic propagation in
2D random liquid media}
\author{Emile Hoskinso\footnote{Address after July 1, 1999:
Department of Physics, University of California, Berkeley,
CA, USA} and Zhen Ye}
\address{Department of Physics and Center for Complex Systems,
National Central University, Chung-li, Taiwan, ROC}

\date{June 10, 1999}

\maketitle

\begin{abstract}

Acoustic wave propagation in liquid media containing many parallel
air-filled cylinders is considered. A self-consistent method is used to
compute rigorously the propagation, incorporating all orders of multiple
scattering. It is shown that under proper conditions, multiple
scattering leads to a peculiar phase transition in acoustic propagation.
When the phase transition occurs, a collective behavior of the cylinders
appears and the acoustic waves are confined in a region of space in the
neighborhood of the transmission source. A novel phase diagram is used to
describe such phase transition.

\end{abstract}

When propagating through media containing many scatterers, waves will be
scattered by each scatterer. The scattered wave will be again scattered by 
other scatterers. Such a process will be repeated to establish an infinite
recursive pattern of multiple scattering, effectively causing the scattering
characteristics of the scatterers to change. Multiple scattering
of waves is responsible for many fascinating phenomena\cite{Ishimaru},
including modulation of ambient sound at ocean surfaces\cite{Vagle},
acoustic scintillation from turbulent flows\cite{DMF}, white paint, 
random lasers\cite{Laser}, electrical resistivity, and photonic band gaps in
periodic structures\cite{Band}. More interesting, perhaps, under proper
conditions multiple scattering leads to the unusual phenomenon of wave
localization, a concept introduced by Anderson\cite{Anderson} to explain the
conductor-insulator transition induced by disorders in electronic systems.
That is, the electrical conductivity can be completely blocked and electrons
remain localized in the neighborhood of the initial emission site due to
multiple scattering of electronic waves by a sufficient amount of impurities
in solids. By analogy, it has been conjectured that similar localization
effect may also exist in the transmission of classical waves in randomly
scattering media.

Considerable efforts have been devoted to propagation of classical waves
in random media. Localization effects have been reported for microwaves in
2D random systems\cite{Microwave}, for acoustic waves in underwater
structures\cite{AC}, and for light\cite{Light}. Research also suggests that
acoustic localization may be observed in bubbly liquids\cite{EPL,PRL}.
Despite the efforts, however, no deeper insight into localization
can be found in the literature, as suggested by
Rusek {\it et al.}\cite{Rusek}. The general cognition is that enhanced
backscattering is a precursor to localization and waves are always
localized in 2D random systems. Important questions such as how
localization occurs and manifests remain unresolved. It is also unknown
how many types of localization exist. It is believed that localization is a
phase transition. However, how to characterize such a phase transition has
not been considered in the literature.  A deeper question may concern whether
wave localization corresponds to a symmetry breaking and whether the
collective behavior often seen in phase transitions such as superconductivity
exits. This paper attempts to shed light on these questions.

In this paper, we present a rigorous study of acoustic propagation
in liquid media containing many air-filled cylinders. The approach is
based upon the self-consistent theory of multiple scattering\cite{Foldy}
and has been used previously to study acoustic localization in bubbly liquid
media\cite{PRL}. Wave propagation is expressed by a set
of coupled equations and is solved rigorously. We report a new phase
transition in acoustic localization in 2D random liquids that not only waves
are confined near the transmitting source but an amazing collective
phenomenon emerges. Unlike previous 2D cases, waves are not always localized.
An essential component of localization in this case arises from the natural
resonance and the collective behavior of the cylinders.

We consider a 2D acoustic system in the cylindrical
coordinates shown in Fig.~1. Assume that $N$ uniform air-cylinders
of radius $a$ are placed randomly in parallel in water perpendicular to
the $x-y$ plane, forming a random cylindrical array. All cylinders are placed
within a circle of radius $R$; the cylinder numerical density is
$N/(\pi R^2)$. The fraction of area
occupied by the cylinders per unit area $\beta$ is $Na^2/R^2$; the average
distance between nearest neighbours is therefore $(\pi/\beta)^{1/2}a$. 
No cylinder is located at the origin, where a line source is placed, and
no two cylinders can occupy the same spot, i.~e. hard sphere approximation.
Experimentally, the air-cylinders can be any gas enclosure with a thin
insignificant elastic layer, like the Albunex bubbles used in
echocardiography\cite{Albunex}. We investigate the acoustic propagation
in such random liquid media.

A unit acoustic line source, transmitting monochromatic pressure waves, is
set at the origin, also perpendicular to the $x-y$ plane. Due to the
large contrast in acoustic impedance between air and water, the air-cylinders
are strong acoustic scatterers. Unlike the spherical bubble case, the
absorption caused by thermal exchange and viscosity is unimportant here and
can be ignored\cite{Weston}.

Multiple scattering in such systems can be computed exactly using the
well-known self-consistent method\cite{Foldy} and the matrix inversion
scheme\cite{PRL}. In the scheme, the multiple scattering can be
solved in terms of the response function of individual scatterers.

The response function of the single cylinder denoted by $\Pi_i$
is readily computed by the modal-series expansion in the cylindrical
coordinates\cite{Hasegawa,YH}; for the present case, $\Pi_1 =
\Pi_2 = \cdots = \Pi_N$. It is found that strong response from a single
cylinder occurs for $ka$ ranging roughly from 0.0005 to 0.5, and a resonant
peak is located at $ka$ around 0.005; here $k$ is the conventional wavenumber
and $a$ is the radius of the cylinders. In this range
of $ka$, to which our attention is restricted, the radial pulsation of the
cylinders, i.~e. the monopole mode, dominates the scattering. In this case,
the scattering from any single cylinder is isotropic in the $x-y$ plane.
The scattering function is plotted against frequency in
Fig.~2, where both real and imaginary parts of the scattering function
are plotted.
The scattering cross section of a single cylinder reaches
a peak. For frequencies above the peak, although it reaches a minimum, 
the scattering function does not vary much.

Consider that the cylinders are located at $\vec{r}_i$ ($i = 1, 2, ..., N$)
in the cylindrical coordinates. The scattered wave $p_s$ from each cylinder,
the $i$-th say, is a linear response to the incident wave composed of the
direct incident wave from the source and {\it all} scattered waves from other
cylinders.

Similar to the 3D case\cite{PRL},
the scattered wave from the $i$-th cylinder is therefore
\BA
p_s(\vec{r}, \vec{r}_i) &=& \Pi_i\left(p_0(\vec{r}_i) +
\sum_{j=1, j\neq i}^{N} p_s(\vec{r}_i, \vec{r}_j)\right)\times
\nonumber\\
& & i\pi H_0^{(1)}(k|\vec{r}-\vec{r}_i|),   \label{eq:1}
\EA
where the second term on the RHS refers to the
multiple scattering, and $p_0$ is the direct incident wave from the source
and equals the usual 2D Green's function for a unit line source, i.~e.
$i\pi H_0^{(1)}(k|\vec{r}|)$; here $H_0^{(1)}$ is the zero-th order Hankel
function of the first kind. To solve for $p_s(\vec{r}_i, \vec{r}_j)$,
we set $\vec{r}$ at one of the scatterers. Then Eq.~(\ref{eq:1})
becomes a set of closed self-consistent equations and can be solved by 
matrix inversion. Once $p_s(\vec{r}_i, \vec{r}_j)$ is determined, the total
wave at any space point is given by
\BE
p(\vec{r}) = p_0(\vec{r}) + \sum_{i=1}^{N}p_s(\vec{r}, \vec{r}_i).
\label{eq:2}
\EE

First consider wave transmission through
the media as a function of frequency.
We plot the ratio $|p(\vec{r})|^2/|p_0(\vec{r})|^2$
in the limit $|\vec{r}| \rightarrow \infty$, giving the
far field transmitted intensity.
A set of numerical computations has been performed for various
area fractions $\beta$, numbers $N$, and dimensionless $ka$.
All parameters are non-dimensionalized.  Distances have units of
the unspecified constant $a$ and functions are dependent only on
the dimensionless parameter $ka$ rather than $k$ and $a$ independently.
Figure 3 presents one of the typical
results for the transmitted intensity in a given direction as a function of
frequency in
terms of $ka$ the air-cylinder distribution shown in Fig.~1.
It is shown that the transmission is significantly reduced
from $ka = 0.006$ to $ka = 0.02$; little energy is transmitted through.
Outside this frequency region, however, there is no significant propagation
inhibition; transmitted intensity varies around the value one.
This result holds for any directions of transmission, i.e. it is representative
of the frequency dependence of the total power escaping to infinity.
In line with the 3D situation\cite{PRL}, the regime in which the
transmission is blocked indicates acoustic localization and is referred
to as the localization regime. When $N$ is increased while keeping $\beta$
constant, the transmitted intensity for $ka$ between 0.006
and 0.02 will decrease as $\exp(-(N/\beta)^{1/2}/(L_a/a))$, from which
the energy localization length $L_a$ can be estimated in terms of the
cylinder radius $a$. Note from Fig.~3 that at certain frequencies,
the transmission is enhanced. Such enhancement can be
due to strong scattering of individual bubbles, multi-reflection at
the sample boundary, or the strong multiple scattering among scatterers
in certain directions, but it is not due to a global collective behavior
of the scatterers. We also note that the localization does
not occur at the resonance peak located at around $ka=0.005$.

To further explore the behavior of acoustic localization in such systems, we
investigate the response of each individual cylinder to the incident waves.
Upon incidence, each cylinder acts effectively as a secondary source. The
scattered wave can be regarded as the sum of the radiated waves from these
secondary sources.
From Eq.~(\ref{eq:1}), the contribution from a given cylinder may be
rewritten as
\BE
p_s(\vec{r}, \vec{r}_i) = i\pi A_i H_0^{(1)}(k|\vec{r}-\vec{r}_i|),
\ (i = 1, 2, ..., N),
\label{eq:3}
\EE
where the complex coefficient $A_i$ denotes the effective strength
of the secondary source and is computed incorporating {\it all} multiple
scattering effects. Note that without the scatterers, obviously we will have
$A_i = \Pi_i p_0(\vec{r}_i) = i\pi \Pi_i H_0^{(1)}(k|\vec{r}_i|)$.

We express $A_i$ as $|A_i|\exp(i\theta_i)$: the modulus $A_i$ represent
the strengths, whereas $\theta_i$ refer to the phase of the effective
secondary sources. We assign a unit vector, $\vec{u}_i$, termed the
phase vector
hereafter, to each phase $\theta_i$, and represent these phase vectors by a
phase diagram in the $x-y$ plane: The phase vectors are written as
$\vec{u}_i = \cos\theta_i\vec{e}_x + \sin\theta_i\vec{e}_y$, where
$\vec{e}_x$ and $\vec{e}_y$ are unit vectors in the $x$ and $y$ directions
respectively; in the phase diagram, the phase vector $\vec{u}_i$ is located
at the cylinder to which the phase $\theta_i$ is associated.

Setting the phase of the initial driving source to zero, numerical experiments
are carried out to study the behavior of the phase vectors and energy spatial
distribution as the incidence frequency varies. We observe an amazing pattern
transition for the phase vectors, with which the wave localization is
correlated. The left column of Fig.~3 shows the phase diagram for the phase
vectors. The right column shows surface plots of the corresponding secondary
source strength magnitudes $|\vec{r}_i||A_i|^2$, giving the energy
distribution scaled to eliminate geometrical spreading effects.
The same cylinder array as in Fig.~1 has been used; three frequencies
below, within, and above the localization regime are chosen according to
the results in Fig.~3.
In this particular case, we have set $\beta = 10^{-3}$ and $N= 200$.

We observe that for frequencies below about $ka = 0.006$\cite{Low}
i.~e. for frequencies on the left side of the localization regime indicated
by Fig.~4, there is no ordering in the phase vectors $\vec{u}_i$.
The phase vectors point to various directions. The energy distribution is
extended in the $x-y$ plane, and no acoustic localization appears.
These are illustrated by the case of $ka = 0.0044$ in Fig.~4.

As the frequency increases, an ordering in the phase vectors and the energy
localization starts to appear. There is a transition period in which the
phase vectors point to either the positive $x$-axis or the negative
$x$-axis, as shown by the case of $ka = 0.008$.
For frequencies fully within the localization regime
indicated by Fig.~3, all phase vectors point to the same direction,
implying that all cylinders oscillates completely in phase. As indicated by
the cases with $ka = 0.012$ and $0.018$ in Fig.~4, all phase vectors point to the
negative $x$-direction, exactly out of phase with the transmitting source.
Such collective behavior
allows for efficient cancellation of incoming waves\cite{YH}.
In this case, the wave energy is localized near the transmitting source.
It decays about exponentially in all directions, setting the localization
length to be around $26a$, much smaller than the range enclosing the random
cylinder array; for
$\beta = 10^{-3}$ and $N = 200$, the range $R$ is around $450a$.
The localization behavior is independent of the outer boundary and always
surfaces for sufficiently large $\beta$ and $N$. In the
field theory, such a collective phenomenon
is an indication of a global behavior of the system and may imply a
symmetry breaking and appearance of a kind of Goldstone boson.

The non-localization at $ka =0.0044$ is not caused by the finite sample size,
because no indication of energy localization appears as we increase the
sample size. If the wave were localized at this frequency, the localization
length would be shorter than that for $ka = 0.012$, as the scattering is
stronger (Fig.~2). The localization would be stronger; this is not observed.

When the frequency increases further, moving near the upper transition
edge, the global in-phase ordering starts to disappear. This is
illstrated by the case of $ka = 0.0216$. When moving out of the localization
regime, the in-phase ordering disappears completely.
In the meantime, the wave becomes
extended again. This is shown by the example $ka = 0.04$ in Fig.~4.

In Fig.~4, we also plot the phase portraits for three frequencies 
($ka = 0.002, 0.0034$ and $0.036$) at
the enhanced transmission shown by Fig.~3. It is shown that there is no
apparent phase ordering in these cases. The figure also show that
the enhancement in one direction does not necessarily imply
enhancement in all directions.

To quantify the phase transitions shown in Fig.~4, we may
define two tentative order parameters. For the in-phase behavior,
the magnitude of the summation of all phase vectors normalized by $N$
may be regarded as a phase order parameter, called
order parameter 1. For the mixed ordering shown
in the transition regime, we may define the order parameter as
$\frac{1}{N}\sum_{i}^N (|u_x(i)| - |u_y(i)|)$, termed order parameter 2.
These order parameters are plotted as a function of frequency in
Fig.~5. When all cylinders oscillate in phase, the order parameter 1 is
close to one, while in the transition regime, the order parameter 2 may
be more proper in describing the phase transition.

The behavior of the system is a result of the interplay between the
single cylinder scatter function and the many body interactions.
Without multiple scattering, the localization and coherence phenomenon
does not occur.  Both the phase and strength of the scattering function
determine how a multiply scattered wave propagates through the medium.
Exactly how the interplay between the scattering function and the multiple
scattering gives rise to the localization and the global coherence behavior
remains an open question and may only be understood by an analytic analysis
which is under our consideration.

The localization or the transmission drop starts at a frequency slightly above the resonant peak, about the minimum of the
real scattering function. Throughout the localization regime, the
scattering function varies mildly. 
The localization or the transmission drop disappears at some frequency which
does not appear to be dependent on the peak or the dip of the
scattering cross section. The phase
transition at the lower frequency edge seems to be related with the small
dip in the scattering cross section or in the imaginary part of scattering 
function, while the transition at the
higher end to the decreasing wavelength. Once the cylinders become
too many wavelengths apart, despite a slight increase in the strength of the
scattering cross section, phase coherence can no longer be maintained
and localization is destroyed.

We have modified the scattering function manually, and observed that
transmission is sensitive to it.  If, however, the scattering function is
held constant at its value for a frequency within the localized range, and
the transmission calculated versus frequency using this constant value,
the phase transitions and a similar transmission behavior are still observed.

The following considerations support and elucidate the localization behaviour
and phase transition observed.
(1) The localization is not due to dissipation.
Acoustic absorption is negligible in the cases considered, and no
mechanism for absorption has been included in our model.
(2) By varying the cylinder number $N$ while keeping the area
fraction $\beta$ constant, it can be shown that the localization or
non-localization behavior is qualitatively unchanged, thus not caused
by the boundary of the cylinder arrays.
(3) The above localization phenomena are caused by multiple scattering.
When we manually turn off the multiple scattering from Eq.~(\ref{eq:1}),
the localization disappears.
(4) The localization behavior also disappears when the cylinder concentration
reaches a sufficiently low level. The localization range in
Fig.~3 is narrowed, tending to vanish.
(5) When the wave is not localized, the energy
distribution varies as the random placement of the cylinders changes at a
given concentration $\beta$. Once
localized, the localization behavior will {\it not change} as the cylinder
placement varies. 

In summary, we have demonstrated a new phase transition in acoustic
propagation in 2D random arrays of air-cylinders in water. When the
concentration of the air-cylinders exceeds a certain
value, acoustic waves become localized near the transmitting source within
a range of frequencies. The results indicate that the wave localization is
related to the collective behaviour of the air-cylinders in the presence of
multiple scattering. 
We take the view that the coherent behavior allows for
efficient cancellation of the source wave, giving rise to localization,
but localization and coherence occur hand-in-hand and
the statement that the localization is due to the coherence or visa-versa
may be just a matter of perspective.
Although these properties may not hold in general,
the fact that they do for resonant air-cylinders makes these scatterers
ideal for theoretical and experimental localization studies.

The work received support from National Science Council of ROC and from
National Central University in the form of a special scholarship to EH.

\section*{Figure Captions}

\begin{description}

\item[Fig. 1]
Top view of a random distribution of air-cylinders. 
The circles refer to the air-cylinders (not to scale).
A straight line source is
placed at the origin.

\item[Fig. 2] The scattering function versus frequency.

\item[Fig. 3] Transmission versus frequency in terms of $ka$.

\item[Fig. 4] Left column: Phase diagram for the two-dimensional phase
vector defined in the text. Right column: Spatial distribution of acoustic
energy (arbitrary scale).

\item[Fig. 5] `Order parameters' versus frequency.

\end{description}


\begin{references}

\bibitem{Ishimaru} A. Ishimaru, {\it Wave propagation and scattering in
random media}, (Academic Press, New York, 1978).

\bibitem{Vagle} D. M. Farmer and S. Vagle,
J. Acoust. Soc. Am. 86, 1897 (1989).

\bibitem{DMF} D. M. Farmer, S. F. Clifford, and J. A. Verall, 
J. Geophys. Res. 92, 5368 (1985).

\bibitem{Laser} N. M. Lawandy, R. M. Balachandran, A. S. L. Gomes, and
E. Sauvain, Nature 368, 436 (1994).

\bibitem{Band} W. M. Robertson, {\it et al.}, 
J. Opt. Soc. Am. B 10, 322 (1993).

\bibitem{Anderson} P. W. Anderson, Phys. Rev. 109, 1492 (1958).

\bibitem{Microwave} R. Dalichaouch, J. P. Amstrong, S. Schultz,
P. M. Platzman, and S. L. McCall, Nature 354,
53 (1991).

\bibitem{AC} C. H. Hodges and J. Woodhouse, J. Acoust. Soc. Am. 74,
894 (1983).

\bibitem{Light} D. S. Wiersma, P. Bartolini, A. Lagendijk, and
R. Roghini, Nature 390, 671 (1997).

\bibitem{EPL} D. Sornette and B. Souillard, Europhys. Lett. 7, 269 (1988).

\bibitem{PRL} Z. Ye and A. Alvarez, Phys. Rev. Lett. 80, 3503 (1998).

\bibitem{Rusek} M. Rusek, A. Orlowski, and J. Mostowski,
Phys. Rev. E. 53, 4122 (1996).

\bibitem{Foldy} L. L. Foldy, Phys. Rev. B67, 107 (1945).

\bibitem{Albunex} N. de Jone and L. Hoff, Ultrasonics 31, 175 (1993).

\bibitem{Weston} D. E. Weston, in {\it Underwater Acoustics}(ed. V. M. Albers)
(Plenum, New York, 1967).

\bibitem{Hasegawa} T. Hasegawa, {\it et al.}, J. Acoust. Soc. Am.
93, 154 (1993);

\bibitem{YH} Z. Ye and E. Hoskinson, {\it Notes on acoustic localization},
unpublished (1998); Z. Ye and E. Hoskinson, {\it Localization of acoustic
propagation in water with air cylinders}, submitted.

\bibitem{Low} For extremely low frequencies our model may not be applicable,
but this is not our interest.

\end{references}
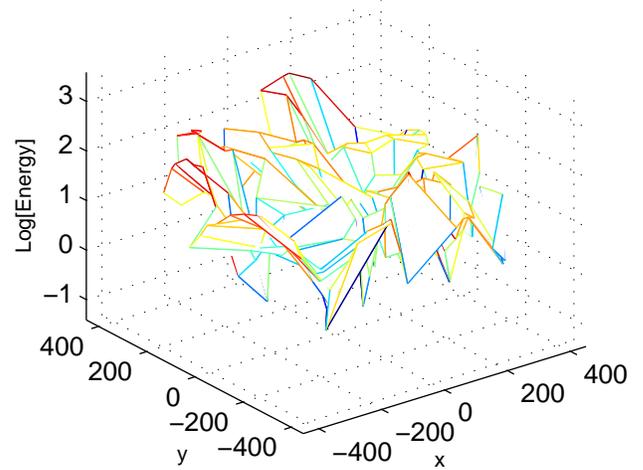
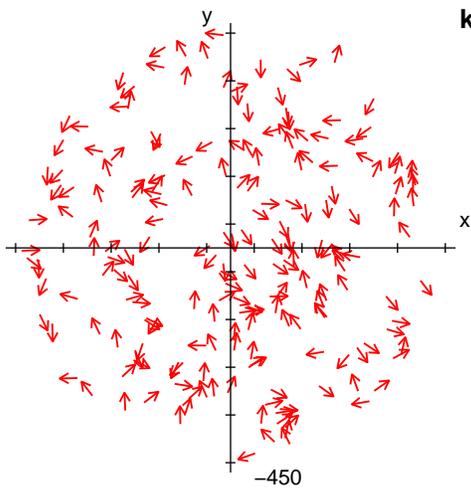
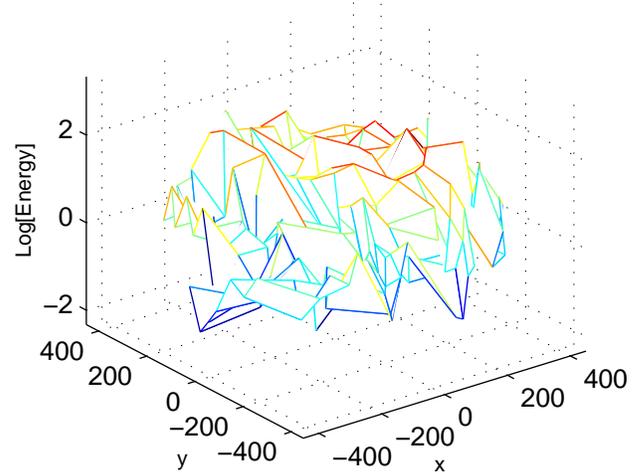
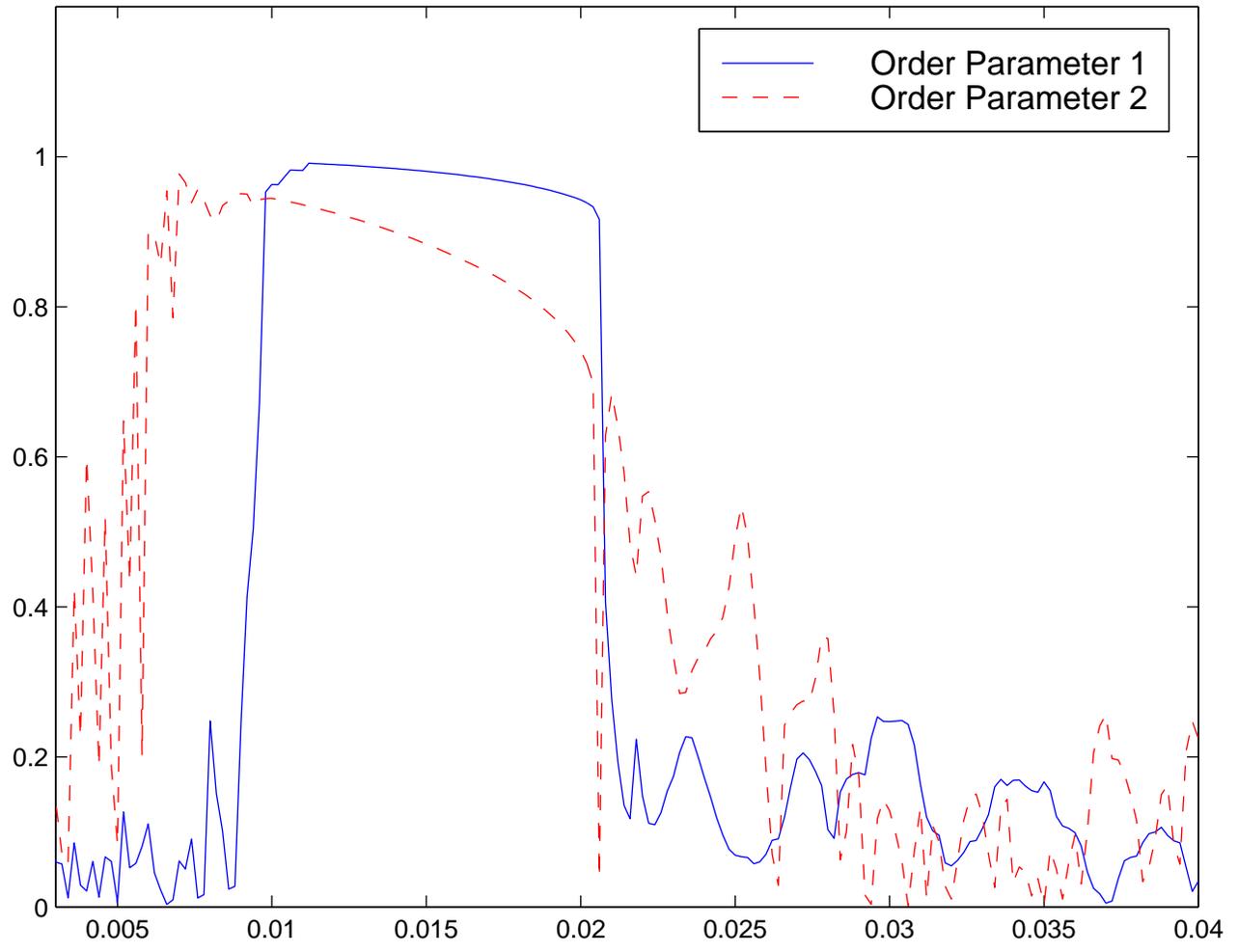
\end{document}